%% file: charm2015_MaurizioMartinelli.tex
\documentclass[12pt]{article}
\usepackage{graphicx}
\usepackage{amsmath}

\newcommand\pubnumber{hep-ex/}
\newcommand\pubdate{\today}

\def\epfl{Laboratoire de Physique des Hautes Energies\\
\'Ecole Polytechnique F\'ed\'erale de Lausanne, CH-1015 Lausanne, Switzerland}

\def\Title#1{\begin{center} {\Large #1 } \end{center}}
\def\Author#1{\begin{center}{ \sc #1} \end{center}}
\def\Address#1{\begin{center}{ \it #1} \end{center}}

\newcommand\pubblock{\rightline{\begin{tabular}{l} \pubnumber\\
         \pubdate  \end{tabular}}}
\newenvironment{Abstract}{\begin{quotation}  }{\end{quotation}}
\newenvironment{Presented}{\begin{quotation} \begin{center} 
             PRESENTED AT\end{center}\bigskip 
      \begin{center}\begin{large}}{\end{large}\end{center} \end{quotation}}

\input econfmacros.tex

\begin{document}
\begin{titlepage}
\pubblock

\vfill
\Title{Measurements of \T-odd observables}
\vfill
\Author{Maurizio Martinelli\\on behalf the LHCb collaboration}
\Address{\epfl}
\vfill
\begin{Abstract}
The study of \T-odd observables using four-body hadronic final states of charm meson decays provides complementary insight to measuring \CP asymmetries via decay rate asymmetries. New results based on the full LHCb dataset are presented.
\end{Abstract}
\vfill
\begin{Presented}
The 7th International Workshop on Charm Physics (CHARM 2015)\\
Detroit, MI, 18-22 May, 2015
\end{Presented}
\vfill
\end{titlepage}
\def\thefootnote{\fnsymbol{footnote}}
\setcounter{footnote}{0}
%

\section{Introduction}

Violation of \CP symmetry in charm meson decays is expected to be extremely small in the Standard Model (SM)~\cite{Bianco:2003vb,Grossman:2006jg}, although recent calculations do not exclude effects up to a few times $10^{-3}$~\cite{Feldmann:2012js,Brod:2011re,Bhattacharya:2012ah}.
This small \CP violation effect in the SM leaves room for beyond-SM effects that, even if small, could produce an asymmetry significantly larger than that predicted from the SM.
The large samples of charm meson decays recorded at LHCb, enable the search for \CP violation at the sub-percent level, hence approaching the largest SM predictions.

Three kinds of search are usually pursued in probing \CP violation. These involve  interference effects between the decay amplitudes, between the mixing and the decay amplitudes, and in the mixing amplitudes~\cite{Bianco:2003vb}.
In the present contribution, an alternative approach using triple-product correlations is shown.

\section{\CP violation and \T-odd correlations}

The search for \CP violation in the charm sector is particularly interesting since the contribution from SM is expected to be very small.
Any large effect can therefore be associated to new particles and phenomena.
In this frame, it is important to study various decay channels and exploit alternative and complementary techniques.

Most of the searches for \CP violation in the charm sector are made with two-body particle decays~\cite{Aaij:2014gsa,Aaij:2013ria}.
While favoured by the very large statistics already collected by the experiments, these channels do not offer the rich variety of interfering contributions from which \CP violation can arise in multi-body decays.
Furthermore, the presence of independent measurable momenta in the final state allows the exploration of new techniques, such as the one using triple-products.

The study of triple-products, and \T-odd observables in particular, offers a different point of view on \CP violation.
This technique has been initially proposed by Valencia~\cite{Valencia:1988it} for quasi-two-body \B mesons decays, but is valid for other hadrons as well.
Valencia observed that a \CP-violating phase difference appears when studying triple-product correlations of decay rates of a particle decay
\begin{equation}
A_B = \frac{\Gamma(k\cdot \epsilon_1\times\epsilon_2 > 0) - \Gamma(k\cdot \epsilon_1\times\epsilon_2 < 0)}{N_B} \propto \sin(\Delta\delta + \Delta\phi),
\end{equation}
where $\Delta\delta$ and $\Delta\phi$ represent the differences of strong and weak phases, respectively.
The observable $A_B$ is therefore sensitive to \CP violation but also to differences in phases introduced by strong interactions.
Since the latter are independent of \CP, one can build a similar asymmetry for the charged-conjugate decay that will only differ in the sign of the weak phases
\begin{equation}
\overline{A}_B \propto \sin(\Delta\delta - \Delta\phi),
\end{equation}
and extract a \CP-violating observables from the combination of the above asymmetries
\begin{equation}
\label{eq:acptodd}
a_{\CP}^{\T-odd} = \frac{1}{2}\left(A_B + \overline{A}_B\right) \propto \cos\Delta\delta\sin\Delta\phi.
\end{equation}
It should be noted that the so-built observable has a complementary feature to the common decay rate asymmetries among charged-conjugate processes
\begin{equation}
a_{\CP}\propto \sin\Delta\delta\sin\Delta\phi.
\end{equation}
While the latter needs a sizeable difference in the interfering amplitudes' strong phases to be sensitive to \CP violation ($\sin\Delta\delta\neq0$), the former has maximum sensitivity when there is not such difference.

The simplest way to define a triple-product \T-odd observable is by using three independent momentum or spin variables from the decay.
In the case of spinless particles in the final state, at least four particles are needed ($A\to abcd$), and the \T-odd observable can be built as a triple-product of three out of the four momenta in the center-of-mass frame of the decay
\begin{equation}
C_T = \vec{p}_{a}\cdot \vec{p}_{b} \times \vec{p}_{c}.
\end{equation}
The two asymmetries
\begin{align}
\nonumber A_T &= \frac{\Gamma(C_T>0) - \Gamma(C_T<0)}{\Gamma}\\ 
\overline{A}_T &= \frac{\overline\Gamma(-\overline{C}_T>0) - \overline\Gamma(-\overline{C}_T<0)}{\overline\Gamma} 
\end{align}
can be measured and combined to extract the \CP-violating observable
\begin{equation}
\atv = \frac{1}{2}\left(A_T - \overline{A}_T\right),
\end{equation}
which differs with respect to Eq.~\ref{eq:acptodd} for a `minus' sign due to a conventional choice in the definition of $\overline{A}_T$.

One advantage of this technique over those comparing the decay rates of charged conjugate decays is that it does not suffer from any flavour-dependent asymmetry.
The asymmetries $A_T$ and $\overline{A}_T$ are indeed measured separately on charged-conjugate decays, and any tagging or particle reconstruction asymmetry is cancelled in the ratio.
Since these effects are the source of the largest systematic uncertainties in the analyses that compare the decay rates of charged conjugate decays, the systematics uncertainties in this technique are usually very small.

\section{Previous searches}
The first attempts at searching for \CP violation by using triple-product correlations in charm decays were made by FOCUS~\cite{Link:2005th} with a few hundreds of events in $\Dz\to\Kp\Km\pip\pim$, $\Dp\to\KS\Kp\pip\pim$ and $\Ds\to\KS\Kp\pip\pim$ decays\footnote{Throughout this document the use of charge conjugate reactions is implied, unless otherwise indicated.}, obtaining a sensitivity ranging from 5\% to 7\%.

The first measurement reaching the sub-percent sensitivity was made by BaBar~\cite{delAmoSanchez:2010xj,Lees:2011dx}, that obtained sensitivities ranging from 0.5\% to 1\% on the same channels with about 50,000 and 20,000 \Dz and $\Dp_{(s)}$ decays, respectively. 
None of the experiments observed significant deviations from zero.

\section{Search at LHCb}

\begin{figure}[htb]
\centering
\includegraphics[width=0.35\textwidth]{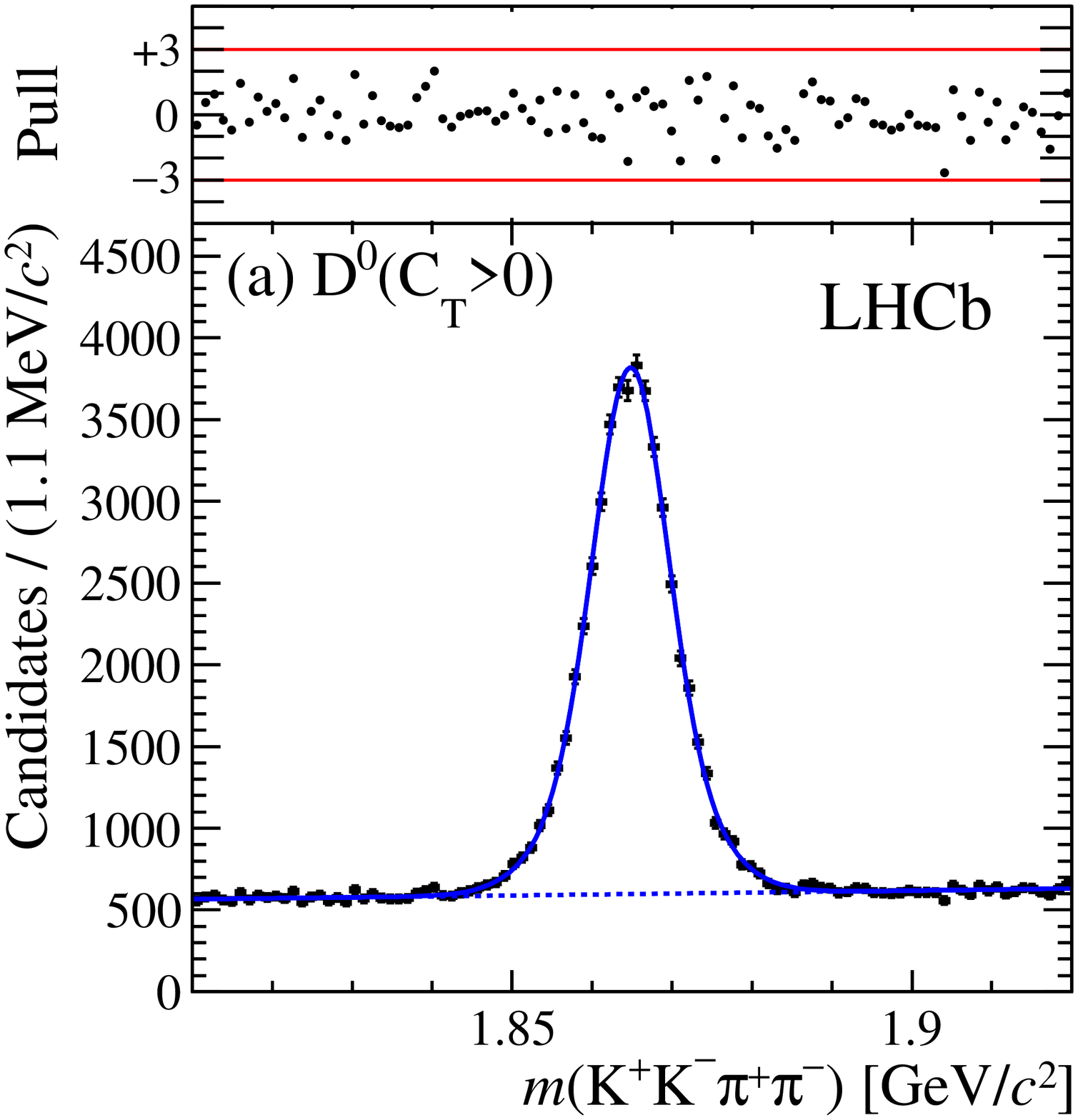}
\includegraphics[width=0.35\textwidth]{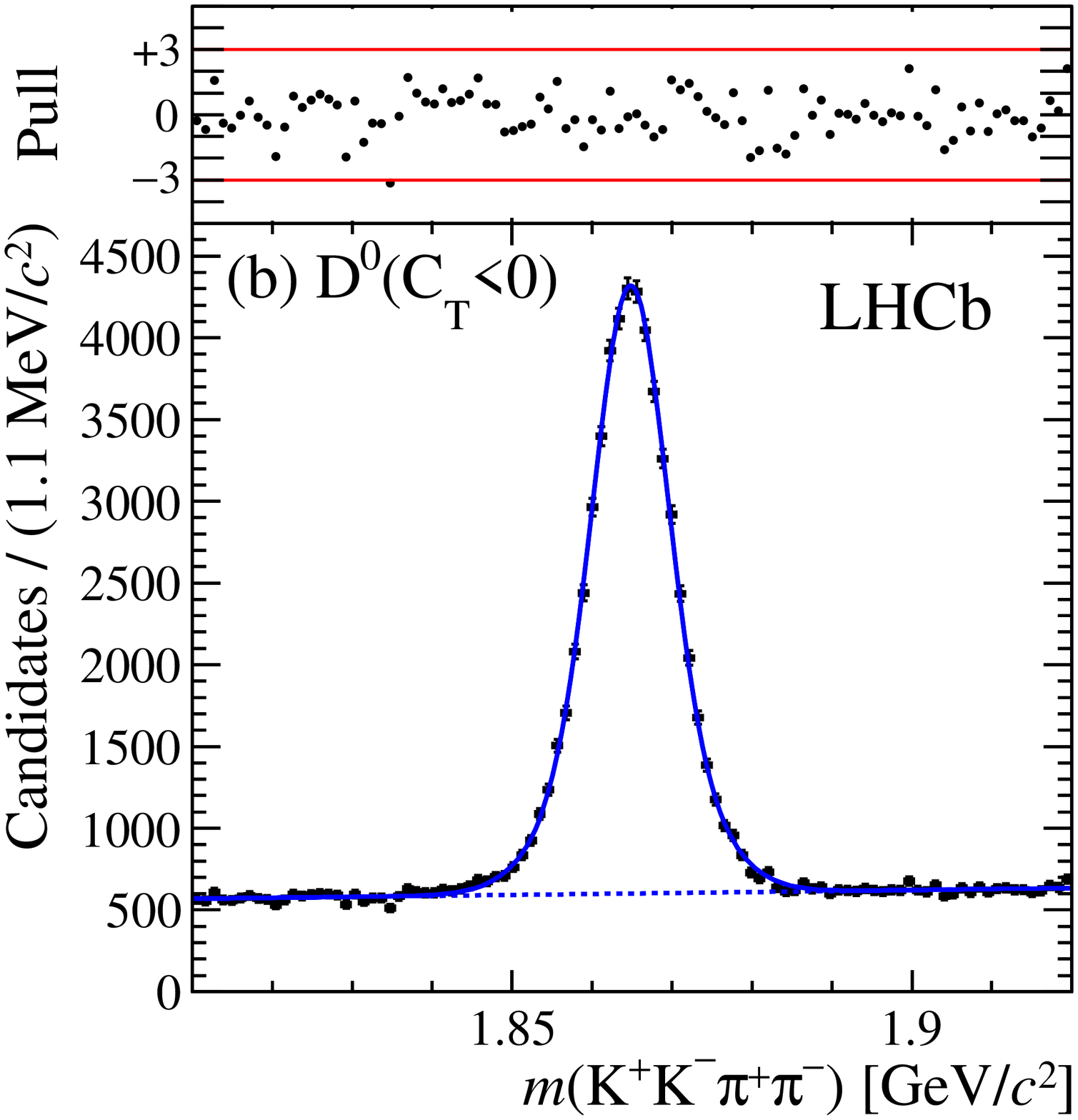}\\
\includegraphics[width=0.35\textwidth]{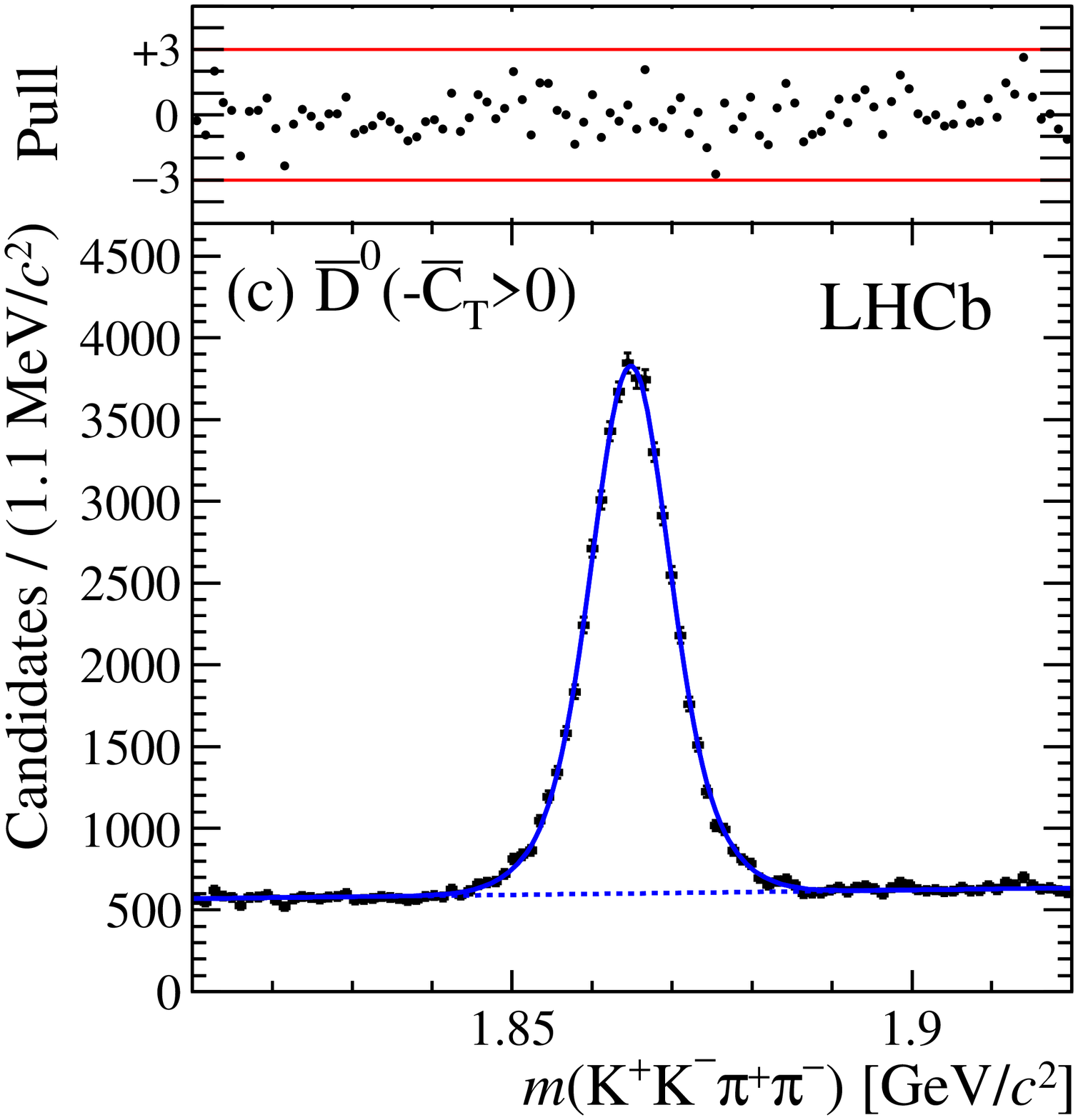}
\includegraphics[width=0.35\textwidth]{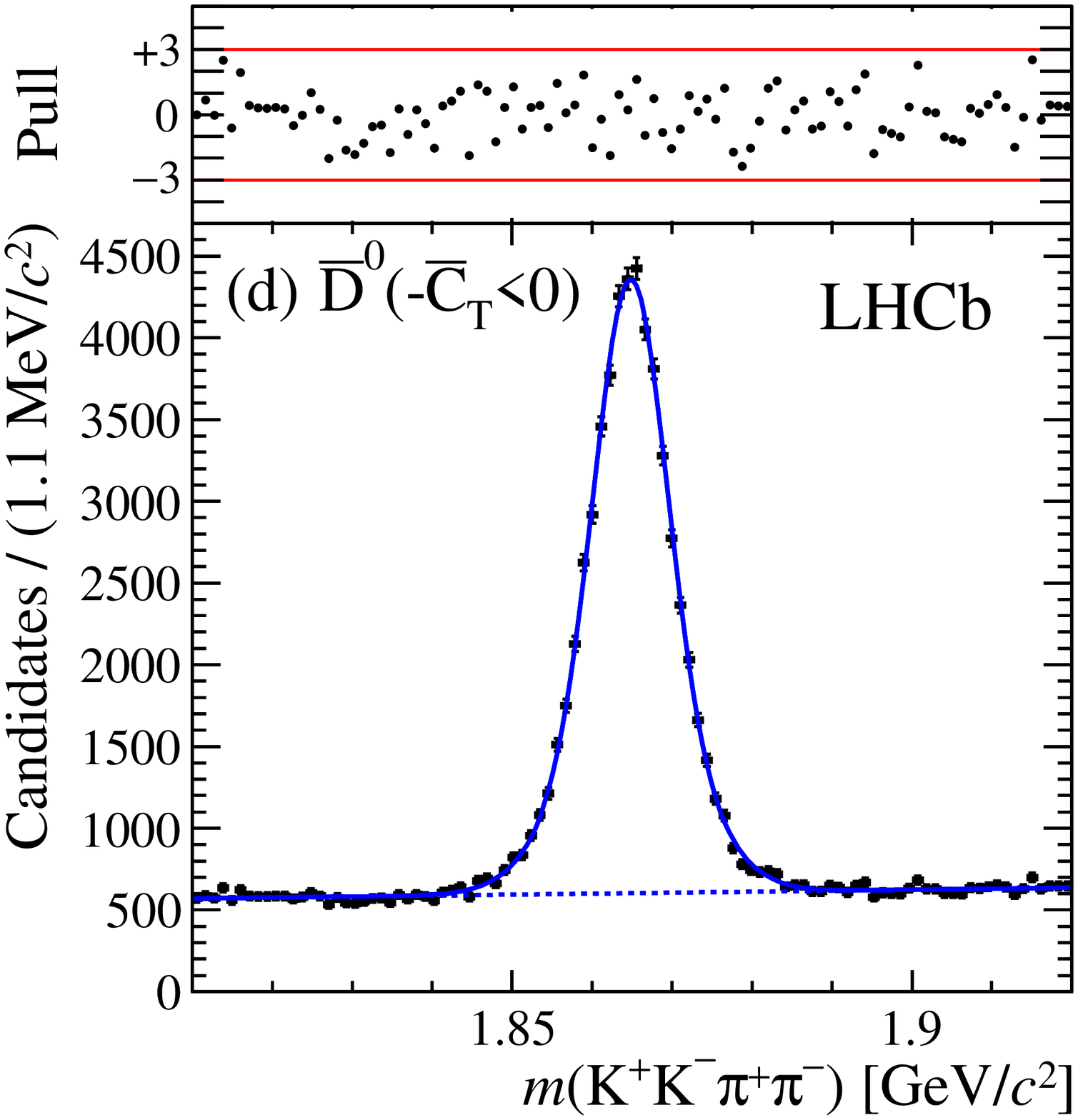}
\caption{Distributions of the $\Kp\Km\pip\pim$ invariant mass in the four samples defined by \Dz (\Dzb) flavour and the sign of $C_T$ ($\overline{C}_T$). 
The results of the fit are overlaid as a solid line, and a dashed line is used for representing the background. 
The normalised residuals (pulls) of the difference between the fit results and the data points, divided by their uncertainties, are shown on top of each distribution.}
\label{fig:d0mass}
\end{figure}

The LHCb experiment searched for \CP violation using \T-odd correlations in $\Dz\to\Kp\Km\pip\pim$ decays~\cite{Aaij:2014qwa}.
The triple-product is defined by using the momenta of three out of the four daughters in the \Dz rest frame ($\ct\equiv\vec{p}_{\Kp}\cdot\vec{p}_{\pip}\times\vec{p}_{\pim}$).
A sample of 170,000 \Dz decays is found in 3 \invfb of data recorded by the LHCb detector in 2011 and 2012, when selecting them through partial reconstruction of semileptonic decays of the \B meson ($\B\to\Dz\mum X$, where $X$ indicates any system composed of charged and neutral particles).
The charge of the muon is used to identify the flavour of the \Dz meson.
The distributions of the \Dz meson candidates in four different regions defined by \Dz flavour and \ct value being greater or less then zero are shown in Figure~\ref{fig:d0mass}.
The four samples shown in the plot are fitted simultaneously to a model made of two Gaussian distributions for signal and and exponential shape for background. 
The asymmetry parameters are extracted directly from the fit.

\begin{figure}[htb]
\centering
\includegraphics[width=0.35\textwidth]{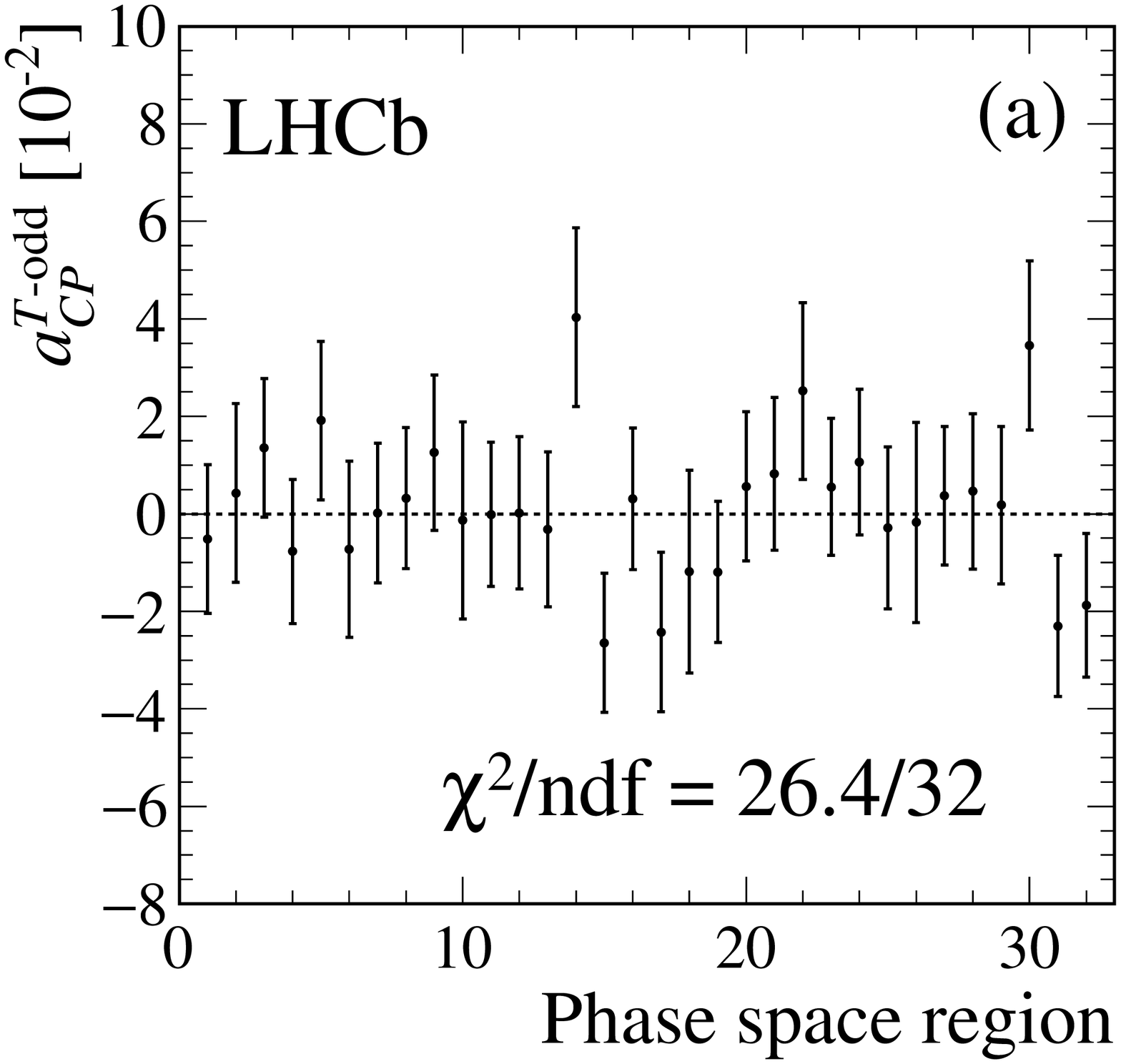}
\includegraphics[width=0.35\textwidth]{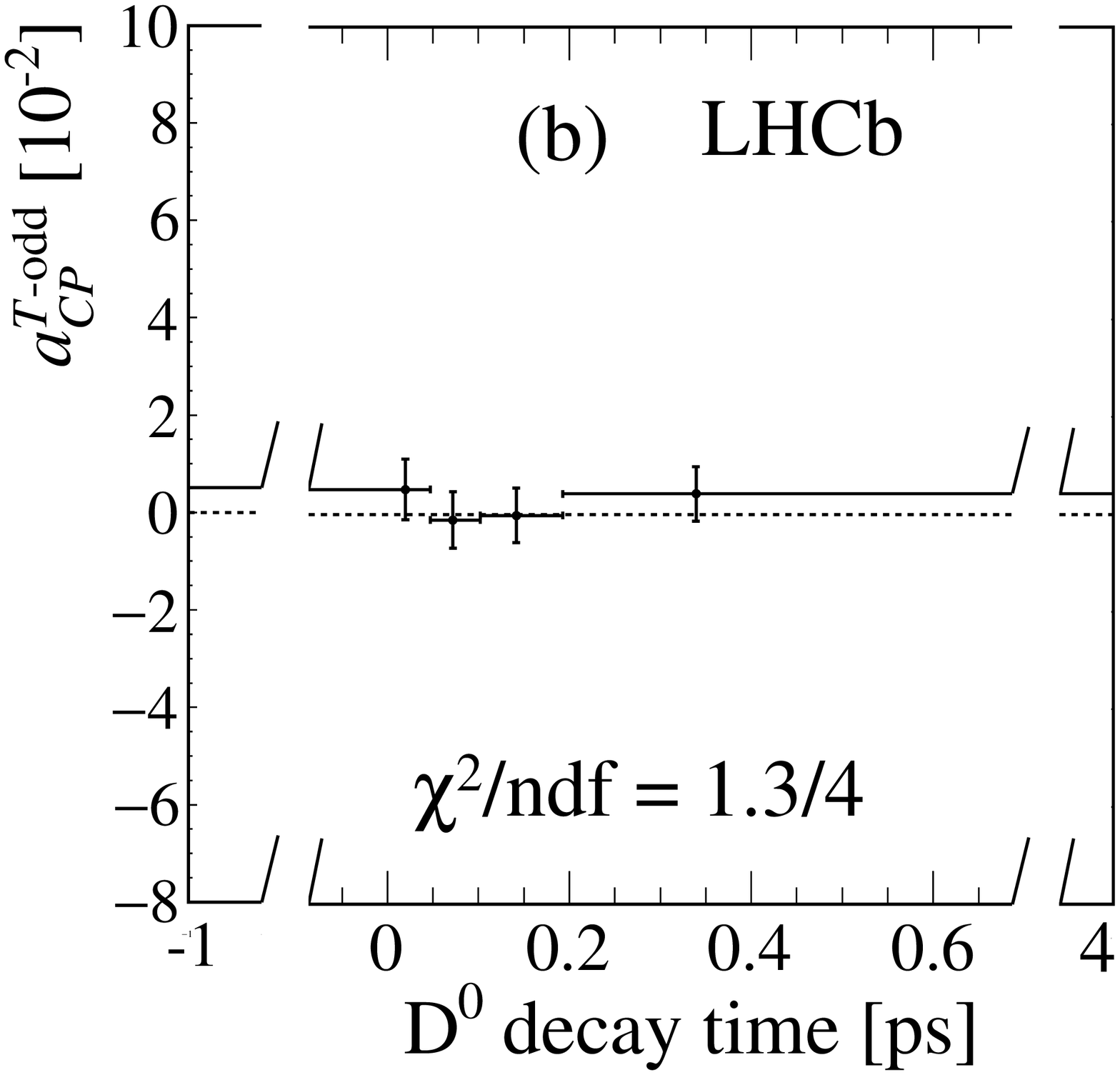}
\caption{Distributions of \atv in (a) 32 bins of the phase space and (b) 4 bins of \Dz decay time.
A $\chi^2$ with respect to the $\atv=0$ hypothesis is calculated and shown in the plot with the relevant number of degrees of freedom.}
\label{fig:atv_bins}
\end{figure}

Three measurements of \CP violation are performed: (i) integrated; (ii) in bins of the phase space; (iii) in bins of decay time.
The phase space is divided in 32 bins defined by means of a Cabibbo-Maksymowicz parametrisation~\cite{Cabibbo:1965zzb}, while 5 bins are used in decay time.
The criterium used to define the bins guarantees a consistent number of events per bin.
The result of the integrated measurement is 
\begin{align}
\atv = (1.8 \pm 2.9(\text{stat}) \pm 0.4(\text{syst})) \times 10^{-3}.
\end{align}
For the binned methods, the asymmetry is calculated in each bin and a $\chi^2$ with respect to the hypothesis of $\atv=0$ is used to estimate the level at which \CP is conserved.
These results are shown in Figure~\ref{fig:atv_bins}.

\begin{figure}[htb]
\centering
\includegraphics[width=0.35\textwidth]{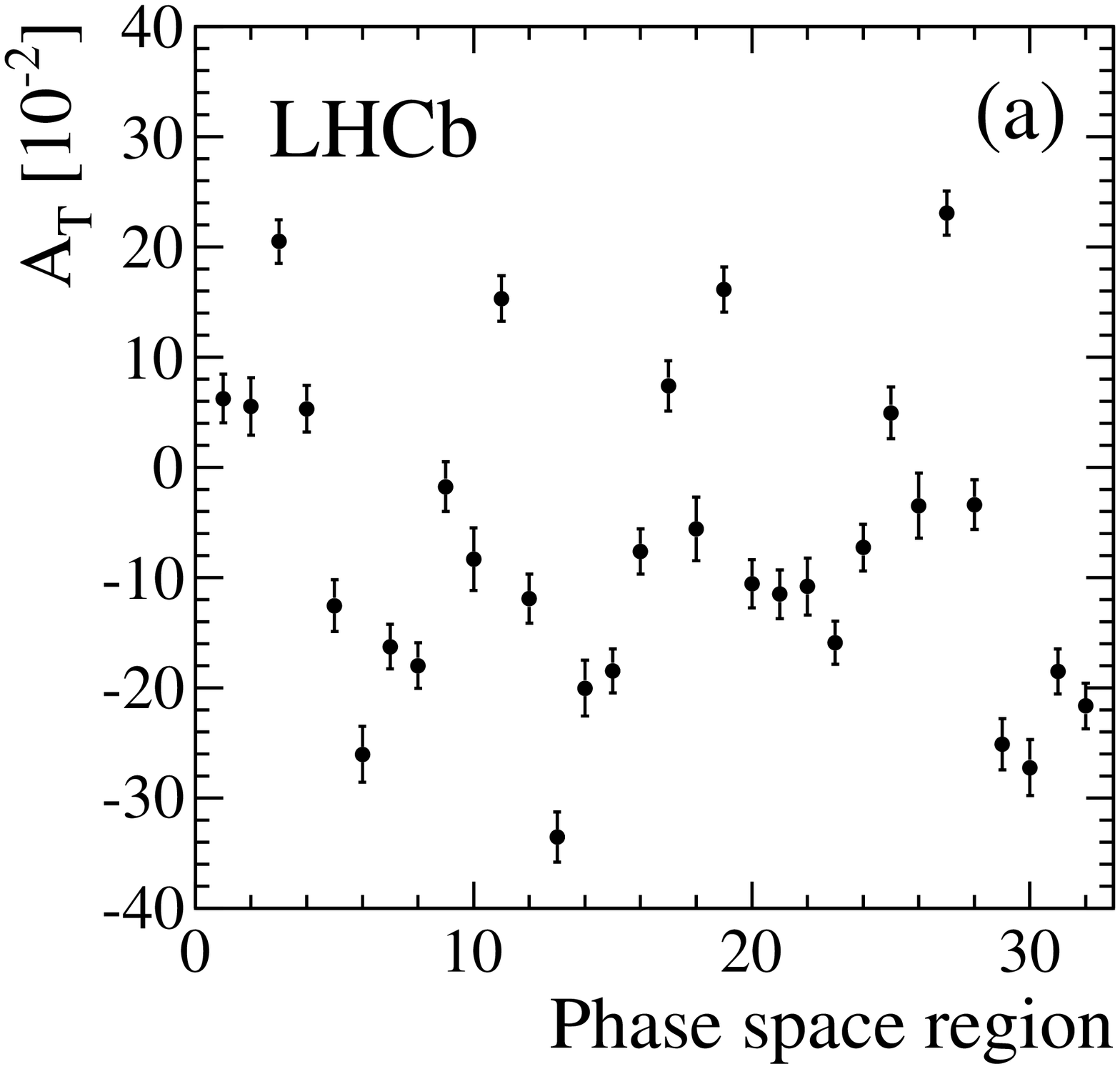}
\includegraphics[width=0.35\textwidth]{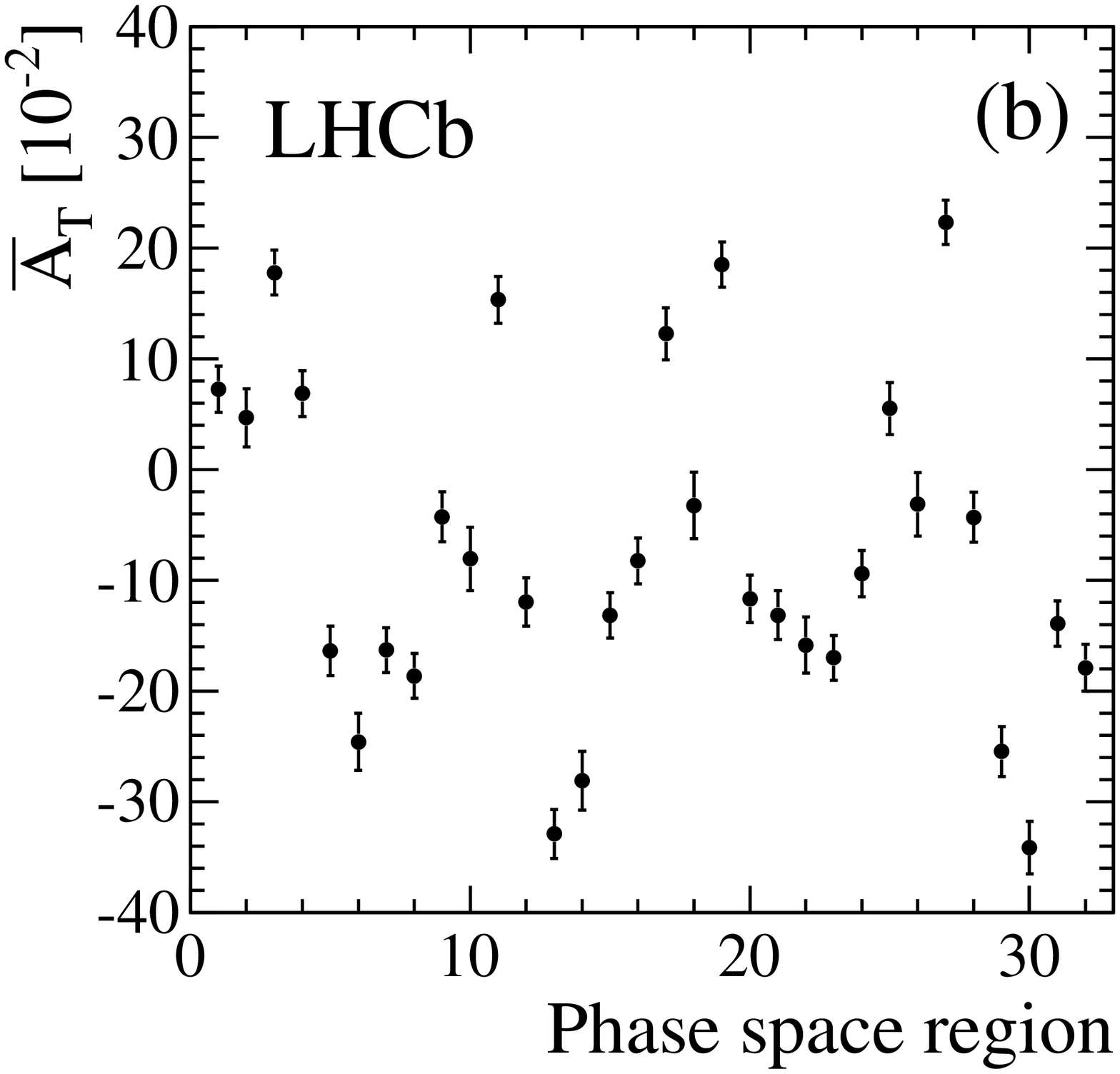}
\caption{Distributions of (a) \at and (b) \atb in 32 bins of the phase space.}
\label{fig:at_bins}
\end{figure}

Useful information can also be extracted by the non-\CP-violating asymmetries \at and \atb.
The integrated measurement shows a significant deviation from zero,
\begin{align}
\at &= (-71.8 \pm 4.1(\text{stat}) \pm 1.3(\text{syst})) \times 10^{-3}\nonumber\\
\atb &= (-75.5 \pm 4.1(\text{stat}) \pm 1.2(\text{syst})) \times 10^{-3}.
\end{align}
While these asymmetries are flat in \Dz decay time, they show significant deviation from the average values in different regions of the phase space, with local asymmetries values ranging from -30\% to 30\%, as shown in Figure~\ref{fig:at_bins}.
These effects can be interpreted as being due to the various resonant contributions that produce different asymmetries as a result of final state interactions in different regions of the phase space.
This is not the only possible explanation though, following a recent interpretation of triple-product asymmetries~\cite{Bevan:2014nva}, part of them can be explained as a signal of \P violation.

An interesting feature of this measurement is that, by definition, the systematic uncertainties are very small.
A detailed study is presented in the LHCb analysis, where the largest uncertainty on \atv is due to detector bias as estimated from a control sample of $\B\to\Dz(\Km\pip\pip\pim)\mum X$ decays. 
This is conservatively assigned as the statistical uncertainty of the measurement on this sample, since no significant bias is observed.
For the \at and \atb measurements, the background from prompt \Dz decays and flavour misidentification are the largest sources of uncertainty, but they cancel in \atv.
Cross-checks were made on particles reconstruction efficiency, particle identification and tagging efficiency, none of them affecting the measurement.

\section{Conclusions}

The triple-product correlations provide alternative and complementary measurements with which to search for \CP violation in multi-body particle decays.
Recent studies suggest that these correlations can be used to probe \C and \P symmetries as well~\cite{Bevan:2014nva}, and that they can be used to probe \CP violation systematically in differential decay distributions~\cite{Durieux:2015zwa}.

The LHCb collaboration searched for \CP violation using \T-odd correlations in $\Dz\to\Kp\Km\pip\pim$ decays, finding $\atv=(1.8\pm2.9(\text{stat})\pm0.4(\text{syst}))\times10^{-3}$.
This result is consistent with the previous ones from BaBar and FOCUS collaborations, and has the best sensitivity so far. 
All of them show no sign of \CP violation.

This observable is affected by systematic uncertainties that are very small, and therefore is appropriate for the study of very large data samples expected at LHCb after LHC Run 2 ($\sim10$ \invfb) or at future experiments, such as Belle-II and the LHCb Upgrade.


\end{document}

%% file: econfmacros.tex
\def\beq{\begin{equation}}
\def\eeq#1{\label{#1}\end{equation}}
\def\eeqn{\end{equation}}

\def\beqa{\begin{eqnarray}}
\def\eeqa#1{\label{#1}\end{eqnarray}}
\def\eeqan{\end{eqnarray}}

\let\bar=\overbar

\def\Dslash{\not{\hbox{\kern-4pt $D$}}}
\def\dslash{\not{\hbox{\kern-2pt $\del$}}}

\def\msb{{\bar{\ssstyle M \kern -1pt S}}}

\RequirePackage{xspace}
\usepackage{relsize}
\def\to{\ensuremath{\rightarrow}\xspace}
\def\CP{\ensuremath{C\!P}\xspace}
\def\C{\ensuremath{C}\xspace}
\def\P{\ensuremath{P}\xspace}
\def\T{\ensuremath{T}\xspace}
\def\B{\ensuremath{B}\xspace}

\def\Dz{\ensuremath{D^0}\xspace}
\def\Dzb{\ensuremath{\bar{D}^0}\xspace}
\def\Dp{\ensuremath{D^+}\xspace}
\def\Ds{\ensuremath{D_s^+}\xspace}
\def\KS{\ensuremath{K^{0}_S}\xspace}

\def\mum{\ensuremath{\mu^-}\xspace}
\def\Kp{\ensuremath{K^+}\xspace}
\def\Km{\ensuremath{K^-}\xspace}
\def\pip{\ensuremath{\pi^+}\xspace}
\def\pim{\ensuremath{\pi^-}\xspace}

\def\ct{\ensuremath{C_T}\xspace}

\def\at{\ensuremath{A_T}\xspace}
\def\atb{\ensuremath{\overline{A}_T}\xspace}
\def\atv{\ensuremath{a_{\CP}^{\T\text{-odd}}}\xspace}

\def\invfb{\ensuremath{\mathrm{fb}^{-1}}\xspace}

\newcommand{\tev}{\ensuremath{{\mathrm{\,Te\kern -0.1em V}}}\xspace}
